\documentclass[11pt,a4paper]{article}

\usepackage{bbding}
\usepackage{pifont}
\usepackage{wrapfig}
\usepackage{amsmath}
\usepackage{verbatim}
\usepackage{amssymb}
\usepackage{inputenc}
\usepackage{textcomp}
\usepackage{appendix}
\usepackage{mathrsfs}
\usepackage{float}
\usepackage{xfrac}
\usepackage{dsfont}
\usepackage{xcolor}
\usepackage{mathtools}
\usepackage{authblk}

\usepackage[
backend=biber,
style=numeric
]{biblatex}
\newcommand{\be}{\begin{equation}}
\newcommand{\ee}{\end{equation}}
\newcommand{\bea}{\begin{eqnarray}}
\newcommand{\eea}{\end{eqnarray}}

\title{Tensionless limit of super-strings and its Majorana condition.}
\author[1,2]{Lucas Delage}
\affil[1]{\small Instituto de Matemática (INSTMAT), Universidad de Talca, Casilla 747, Talca, Chile}
\affil[3]{\small Centro de Estudios Científicos (CECs), Arturo Prat 514, Valdivia, Chile.}

\begin{document}
\maketitle

\begin{abstract}
    We compute the deformation of the spinor fields of the super-Polyakov action leading to the so-called tensionless super-string theory. This allows us to obtain a satisfying Majorana condition in the tensionless limit, which is the main result of this paper. Our tensionless limit is shown to possess previously computed features of tensionless super-string theory.   
\end{abstract}

\section{Introdution}

\paragraph{} In order to understand a particular phenomena, it may be useful to look at some extremal regime of it. It can give some insight to solve the full problem and may lead to simplification hypothesis. It is also often close to the real physical situation, but yet much simpler to analyze. For example in chemistry, the thermodynamic limit, obtained after letting time to go to infinity, informs us of what will be the final concentrations of the different products of the reaction you are considering, whereas the kinetic limit tells how these concentration evolve at the early stage of the reaction. Another, and very famous example, is the black body radiation, for which the Rayleigh-Jean's and Wien's laws, describing respectively the radiation in the limits of long and short wavelength, were found before Max Planck solved the entire problem in his most famous work. Last but not least, the massless limits of quantum fields theories give very satisfactory approximation of high energy regime in particle physics. In the present paper we will also be interested in such extremal limit in the context of tensionless strings, which is nowadays an active subject of research.  

\paragraph{}
The first approach to tensionless strings was provided by Schild \cite{Sch}. Its interest has increased to concern string scattering \cite{GM1}, \cite{GM2}, AdS/CFT correspondence \cite{Kni} or even in Hagedorn phase transition \cite{AtWi}, 
\cite{PiAl}, \cite{Ole}. Since supersymmetry is a central element in string theories, there have been also some works regarding super-string tensionless limits; with the precursor work of \cite{LiSu}. More recently, tensionless limit of the super-Polyakov action have been investigated in which the spinor fields scale inhomogeneously \cite{Ba1}, \cite{Ba2}, \cite{Ba3}. In these works a new action have been obtained, in which the spinor fields play a more important role. Nevertheless, it was not found any Majorana condition in their tensionless action. This is somewhat intriguing since the spinors of the super-Polyakov action are indeed Majorana spinors.

\paragraph{}
Let us remind the expression of the super-Polyakov action: 
\be{} \label{action} 
S=-\frac{1}{4 \pi  \ell}\int d^2 x \sqrt{-g} g^{\mu \nu} \big[\partial_\mu X \partial_\nu X + i\bar{\psi} \gamma_\mu\partial_\nu \psi\big].
\ee
The metric $g$ is the standard Minkowski metric, diagonal in the system of coordinates $(x,t)$, with $g_{xx} = 1$ and $g_{tt} = -1$. The spinor $\psi$ is a two-components Majorana spinor, $\bar{\psi}$ stand for its Majorana conjugate, $\ell$ is the string length. We consider closed strings, with periodicity $T$ along the $x$-coordinates: $X(x+T,t) = X(x,t)$, $\psi(x+T,t) = \psi(x,t)$. Usually, the physical space is supposed to have several dimensions, implying that the fields $X$ and $\psi$ possess several coordinates (i.e. there are collections of fields $\{X^\mu\}$, $\{\psi^\mu\}$). In order to keep the present work as simple as possible, and because it doesn't play any role in the problem we are studying at the present time, we do not consider these coordinates and assume the physical world has dimension 1.

\paragraph{}
Considering only the bosonic part of this action
\be{}
S_{\text{Bos}}=-\frac{1}{4 \pi  \ell}\int d^2 x \sqrt{-g} g^{\mu \nu} \partial_\mu X \partial_\nu X,
\ee 
its tensionless limit is obtained after setting
\be{}
\ell \to \infty, \qquad \frac{t}{x} \to 0.
\ee{}
An usual way of implementing this limit it is to introduce a parameter $\lambda \in [0,1]$, producing the deformation 
\be{}
\ell \mapsto \frac{\ell}{\lambda}, \qquad t \mapsto \lambda t,
\ee{}
and then taking the limit $\lambda \to 0$.

\paragraph{} When the full superstring action (\ref{action}) is considered, a deformation of the spinor $\psi$ is also necessary (it can be inferred by analyzing the physical dimension of the spinor). In the past (\cite{Ba2}, \cite{Ba1}, \cite{Ba3}), this deformation has been guessed and a quite interesting tensionless limit has been obtained. However, the authors of the aforementioned articles have not been able to find any Majorana condition fulfilled by the spinor fields in their tensionless limit.  We will show in this article that this is mainly due to the election of their deformation. In order to circumvent this problem, we will show how to compute the appropriate deformation for the spinor, leading to a satisfying Majorana condition, by a method we will now shortly resume. First, we observe that in the bosonic case, we can, instead of deforming the coordinate, and in a manner totally equivalent, deform the metric. Then, in the full superstring action, we will use this deformation of the metric to obtain a deformation of the Clifford algebra. Finally, invoking the fact that, similarly to the bosonic case, a deformation of the metric should be equivalent to a deformation of both the coordinates and the spinor, we transfer the deformation of the Clifford algebra to its irreducible representation. 

\paragraph{} The organization of the paper is as follow. First we briefly review the standard super-string theory, which allows us to introduce our notations and conventions. Then we present the exotic version of the super-string action, from which the tensionless limit will be taken. The differences at the quantum level, which are the significant discrepancies between the two theories are only evoked. In this section, we also show that a choice of convention for the Clifford algebra, namely the convention that the anticommutator $\{\gamma^\mu,\gamma^\nu\}$ is either plus (which is the standard physicist's convention) or minus (which is the standard mathematician's convention) $2 g^{\mu \nu} \mathds{1}$ leads to the two aforementioned theories. Section 4 is devoted to the deformation of the exotic super-string action and its tensionless limit.

\section{Short review of superstrings}

\subsection{Symmetries}

It is well known that the action (\ref{action}) is invariant under a set of transformation called "extended diffeomorphisms" (see \cite{BLT} for an explanation of the terminology), given by
\bea \label{diffeo_1}
\delta_\xi X &=& \xi^\rho \partial_\rho X, \\
\delta_\xi \psi &=& \frac{1}{2} \xi^\rho \partial_\rho \psi - \frac{1}{2} \varepsilon^\lambda{}_\rho \xi^\rho \partial_\lambda \tilde{\gamma} \psi + \frac{1}{4} \partial_\rho \xi^\rho \psi - \frac{1}{4} \varepsilon^\mu{}_\rho \partial_\mu \xi^\rho \tilde{\gamma} \psi, \label{diffeo_2}
\eea
with $\xi$ subject to 
\be \label{restriction_for_xi}
g^{\mu\nu} \partial_\rho \xi^\rho - g^{\mu \rho} \partial_\rho \xi^\nu - g^{\nu \rho} \partial_\rho \xi^\mu = 0.
\ee
The action is also invariant under the supersymmetry transformations given by
\bea
\delta_\epsilon X &=& \bar{\epsilon} \psi , \label{supersymmetry_1} \\ 
\delta_\epsilon \psi &=& -i \partial_\mu X \gamma^\mu \epsilon,
\eea \label{supersymmetry_2}
where the spinor $\epsilon$ is subject to the condition 
\be \label{restriction_for_epsilon}
\gamma^\nu \gamma^\mu \partial_\nu \epsilon = 0.
\ee 




\subsection{Super-field formalism}

The bosonic symmetry transformations acting on the bosonic field can be represented by a differential operator, as it can be seen from (\ref{diffeo_1}). We would like to extend this feature to the whole superalgebra, and represent every (super-)symmetry transformation by a super-differential operator. For this reason, we introduce on-shell super-fields
\be
Y = X + i \bar{\theta} \psi, 
\ee
where $X$ and $\psi$ are on-shell fields and $\theta$ is called a super-coordinate. Here, the use of on-shell fields is required by the fact that the super-symmetry algebra only closes on-shell. In order to work with off-shell fields, it is necessary to introduce the  so-called auxiliary fields. The advantage of this approach is that the addition of such fields to the theory would allow to work with an off-shell closing supersymmetry algebra. In the super-field approach, the auxiliary fields are components along $\bar{\theta} \theta$. The vanishing of this component is also part of the condition that an on-shell super-field fulfills. For more details on this topic, we refer to \cite{DeFr}. See also \cite{CCF} for a both rigorous and comprehensive presentation of the mathematical nature of super-coordinates.

\paragraph{}
We define the following projectors
\bea 
h_\pm^\mu{}_\nu &=& \frac{1}{2}(g^\mu{}_\nu \mp  \varepsilon^\mu{}_\nu), \\
\text{P}_\pm &=& \frac{1}{2} (\mathds{1} \pm \tilde{\gamma}),
\eea 
where 
\be 
\tilde{\gamma} = \frac{1}{2} \epsilon_{\mu \nu} \gamma^\mu \gamma^\nu,
\ee
as well as as the quantities
\bea 
\psi_\pm &=& \text{P}_\pm \psi, \label{def_psi_+}\\
\partial_{\pm \mu} &=& h_\pm^\nu{}_\mu \partial_\nu, \\
\xi_\pm^\mu &=& h_\pm^\mu{}_\nu \xi^\nu, \label{def_xi_+}\\
\bar{\epsilon}_\pm &=& \bar{\epsilon} \text{P}_\pm = \overline{\epsilon_\mp}.
\eea
These definitions allow a clear decomposition of the symmetries. 

\paragraph{}
Let us show how the super-differential operator representation s obtained by taking the example of the symmetry generated by $\xi_+$. Its action on $Y$ is given by
\be 
\delta_{\xi_+} Y = \delta_{\xi_+} X + i \bar{\theta} \delta_{\xi_+} \psi.
\ee 
We demand it to be of the following form
\be
\delta Y = \delta x^\mu \partial_\mu Y + \delta \bar{\theta}_\alpha \partial_{\bar{\theta}_\alpha} Y.
\ee
Thus, we find expressions $\delta x^\mu ({\xi_+})\partial_\mu$, $\delta \theta_\alpha ({\xi_+})\partial_{\bar{\theta}_\alpha}$ and write
that we call the super-differential operator representation of $\delta_{\xi_+}$. Concerning the notations, we will write this representation as $\text{L}_\pm(\xi_\pm)$, $\text{Q}_\pm(\bar{\epsilon}_\pm)$. Explicitly, we have
\bea \label{L_xi_+}
\text{L}_+(\xi_+) &=&  \xi_+^\mu \partial_{+\mu} + \frac{1}{2} \partial_{+\mu} \xi^\mu_+   \bar{\theta}_+ \partial_{\bar{\theta}_+}, \\
\text{L}_-(\xi_-) &=&  \xi_-^\mu \partial_{-\mu} + \frac{1}{2} \partial_{-\mu} \xi^\mu_-   \bar{\theta}_- \partial_{\bar{\theta}_-}, \label{L_xi_-}\\
\text{Q}_+(\bar{\epsilon}_+)&=& \bar{\theta}_+ \gamma^\mu \epsilon_- \partial_{+\mu} -i\bar{\epsilon}_+  \partial_{\bar{\theta}_+}, \\
\text{Q}_-(\bar{\epsilon}_-) &=& \bar{\theta}_- \gamma^\mu \epsilon_+ \partial_{-\mu} -i\bar{\epsilon}_-  \partial_{\bar{\theta}_-}.
\eea

\subsection{Mode expansion}

The analysis of tensile superstring is simplified by choosing an appropriate coordinate system and an appropriate representation of the Clifford algebra. We use the light cone coordinates: $x^+ = t + x$, $x^- = t - x$, and
\bea 
\gamma_0 &=& \begin{pmatrix} 0 & 1 \\ -1 & 0 \end{pmatrix}, \\
\gamma_1 &=& \begin{pmatrix} 0 & 1 \\ 1 & 0 \end{pmatrix}, \\
C &=& D = \gamma_0, \\
\psi &=& \begin{pmatrix} \psi_- \\ \psi_+ \end{pmatrix},
\eea
where, with a slight abuse of notation, we have identified $\psi_+$ as defined in (\ref{def_psi_+}) and its only one non-vanishing component. 
Using this convention, the equations of motions are 
\bea
&\partial_+ \partial_- X = 0,& \\ \label{eom_X}
&\partial_- \psi_+ = 0, \qquad \partial_+ \psi_- = 0.& \label{eom_psi}
\eea
The solutions of these equations can be expanded in Fourier modes as
\bea
&X = C_0 + \frac{2 \pi \ell}{T} P_0 t + \frac{i \sqrt{\ell}}{2 \pi} \sum\limits_{n \in \mathds{Z}\setminus\{0\}} \displaystyle{ \frac{1}{n} \left( a_n e^{-\frac{2 \pi i n x^+}{T}} + \tilde{a}_n e^{-\frac{2 \pi i n x^-}{T}} \right)},& \label{mode_exp_of_X}\\
&\psi_+(x^+) = \sqrt{2 \pi \ell} \sum\limits_{n \in \mathds{Z}} b^+_n e^{-\frac{2 \pi i n x^+}{T}},& \label{mode_exp_of_psi+} \\
&\psi_-(x^-) = \sqrt{2 \pi \ell} \sum\limits_{n \in \mathds{Z}} b^-_n e^{-\frac{2 \pi i n x^-}{T}}.& \label{mode_exp_of_psi-}
\eea

The symmetry parameters can also be expanded in Fourier modes (the symmetries have to preserve the periodicity of the fields)
\bea 
\xi^+(x^+) &=& \sum_{n \in \mathds{Z}} \xi^+_n e^{-\frac{2 \pi i n x^+}{T}}, \qquad \xi^-(x^-) = \sum_{n \in \mathds{Z}} \xi^-_n e^{-\frac{2 \pi i n x^-}{T}}, \label{mode_exp_of_xi} \\
\bar{\epsilon}_+ (x^+) &=& \sum_{n \in \mathds{Z}} \bar{\epsilon}_{+,n} e^{-\frac{2 \pi i n x^+}{T}}, \qquad \bar{\epsilon}_- (x^-) = \sum_{n \in \mathds{Z}} \bar{\epsilon}_{-,n} e^{-\frac{2 \pi i n x^-}{T}}. \label{mode_exp_of_epsilon}
\eea 
In the light cone coordinate system, we can identify $\xi^+$ (the component of $\xi$ along $x^+$) with $\xi_+$ (defined in (\ref{def_xi_+})). We can therefore perform a mode expansion of the operators $\text{L}_\pm$, $\text{Q}_\pm$ of the previous section. We write
\be
\text{L}_\pm (\xi_\pm) = \sum_{n \in \mathds{Z}} \xi^\pm_n \text{L}_{\pm,n}, \qquad \text{Q}_\pm (\bar{\epsilon}_\pm)  = \sum_{n \in \mathds{Z}} \bar{\epsilon}_{\pm,n} \text{Q}_{\pm,n}.
\ee 

Using their explicit representation given in the appendix, we check that  the operators $\{\text{L}_{+,n}, \text{Q}_{+,n}\}$ and $\{\text{L}_{-,n}, \text{Q}_{-,n}\}$ form two independent copies of the super-Witt algebra, whose non vanishing commutation relations are given by:
\bea \label{Svira1}
[\text{L}_{+,n}, \text{L}_{+,m}] &=&  \frac{2 \pi i}{T} (m-n) \text{L}_{+,n+m}, \\
{}[\text{L}_{+,n}, \text{Q}_{+,r}] &=& \frac{2 \pi i}{T} (r- \frac{n}{2}) Q_{+,n+r}, \\
\{\text{Q}_{+,r}, \text{Q}_{+,s}\} &=&  4 i \text{L}_{+,r+s}. \label{Svira2}
\eea{}

\section{The exotic super-string action.}

\subsection{An exotic convention}

The tensionless super-string action is not obtained as a limit of the standard super-string action, but instead from an exotic action, as was pointed out in \cite{Ba1}. We briefly review this result and refer to the quoted article for more details. The exotic super-string action is obtained after replacing the traditional physicist's definition for the Clifford algebra 
\be 
\gamma_\mu \gamma_\nu + \gamma_\nu \gamma_\mu = 2 g_{\mu\nu}, \label{standard}
\ee 
by another choice, preferred by mathematicians
\be 
\gamma_\mu \gamma_\nu + \gamma_\nu \gamma_\mu = - 2 g_{\mu\nu}. \label{exotic}
\ee 


This conventions has repercussion on the so-called Majorana representation of spinors. In full generality, the Majorana conjugation of a spinor $\psi$ is 
\be
\psi^c = \psi^T C,
\ee 
where $C$, called conjugation matrix, is any matrix satisfying 
\be
C \gamma_\mu C^{-1} = \pm \gamma_\mu{}^T, \label{defining_relation_for_C}\\
\ee
In spacetime of dimension $2$, the two choices are possible: there exists two matrices, let us call them $C_+$ and $C_-$ depending on the sign of (\ref{defining_relation_for_C}), that can be used to define the Majorana conjugation. Furthermore, it can be shown that these matrices can be chosen so that they satisfy
\be 
C_\pm{}^T = \pm C^T.
\ee 
However the on-shell closure of the supersymmetry algebra requires to take $C = C_-$. Therefore there is no choice for the super-Polyakov action. Furthermore, the Majorana condition in nothing else than a reality condition. To keep things as clear as possible, we ask $C$, $\gamma_0$, $\gamma_1$ to be represented by real matrices, implying that a Majorana spinor is simply a real spinor ($\psi^\ast$ = $\psi$). With all these considerations understood, when the convention (\ref{standard}) is chosen, and for real representations, it is possible to have $ \gamma_0$ antisymmetric and thus $C = \gamma_0$. On the other hand, when the convention (\ref{exotic}) is chosen, it is impossible to have both $\gamma_0$ real and symmetric, hence we cannot take $C = \gamma_0$.

\paragraph{}
Now, the main physical implication of such a result comes from the Dirac bracket of the spinor field
\be  \label{bracket_of_spinors}
\{\psi^\alpha, \psi^\beta\}_{\text{D.B.}} = -2 \pi i \ell\left( (C \gamma^0)^{-1}\right)^{\alpha \beta}.
\ee 
The consequence is that in the case of the standard super-Polyakov action, i.e. when convention (\ref{standard}) is chosen for the Clifford algebra, the fermionic modes, after canonical quantization, will follow the anticommutation relations of (infinitely many) standard fermionic harmonic oscillators,
\be \label{harmonic}
[\frac{1}{\sqrt{n}}\hat{b}^+_n, \frac{1}{\sqrt{n}} \hat{b}^+_n{}^\dagger]_+ = 1.
\ee 
On the other hand, in the case of the exotic super-Polyakov action, i.e. when the convention (\ref{exotic}) is used, one of the modes, $\hat{b}^+_n$ say, will follow (\ref{harmonic}) 
whereas the other , $\hat{b}^-_n$ say, will follow 
\be 
[\frac{1}{\sqrt{n}}\hat{b}^-_n, \frac{1}{\sqrt{n}}\hat{b}^-_n{}^\dagger]_+ = -1.
\ee
This kind of commutation relations leads to negative norm states. Differences between standard fermionic oscillators and this kind of exotic fermionic oscillator are surveyed for example in \cite{HeTe}.

\subsection{Symmetries of the exotic super-string action}

The appearance of the minus sign in the right-hand-side of (\ref{exotic}) has repercussion in the (super-)symmetry transformations, which become
\bea
\delta_\xi X &=& \xi^\rho \partial_\rho X, \\
\delta_\xi \psi &=& \frac{1}{2} \xi^\rho \partial_\rho \psi + \frac{1}{2} \varepsilon^\lambda{}_\rho \xi^\rho \partial_\lambda \tilde{\gamma} \psi + \frac{1}{4} \partial_\rho \xi^\rho \psi + \frac{1}{4} \varepsilon^\mu{}_\rho \partial_\mu \xi^\rho \tilde{\gamma} \psi, \\
\delta_\epsilon X &=& -\bar{\epsilon} \psi, \\ 
\delta_\epsilon \psi &=& -i \partial_\mu X \gamma^\mu \epsilon.
\eea 
The restrictions for the (super-)parameters (\ref{restriction_for_xi}), (\ref{restriction_for_epsilon}), however, are unaffected. Because we want that a clear separation (on-shell) between the "+" symmetries and the "-" ones (for example we want $\{\text{Q}_+, \text{Q}_+\} = \text{L}_+$), we have to change the definition of the projectors. Now we set
\be \label{other_projectors}
\text{P}_+ = \frac{1 - \tilde{\gamma}}{2}, \qquad \text{P}_- = \frac{1 + \tilde{\gamma}}{2}.
\ee


The super-differential operators representation of the bosonic symmetries $\text{L}_\pm(\xi_\pm)$ is still given by (\ref{L_xi_+}-\ref{L_xi_-}) (taking (\ref{other_projectors}) into account), whereas for the fermionic symmetries it changes as
\bea
\text{Q}_+(\bar{\epsilon}_+)  &=& \bar{\theta}_+ \gamma^\mu \epsilon_- \partial_{+\mu}  + i\bar{\epsilon}_+ \partial_{\bar{\theta}_+}, \\
\text{Q}_-(\bar{\epsilon}_-)  &=& \bar{\theta}_- \gamma^\mu \epsilon_+ \partial_{+\mu}  + i\bar{\epsilon}_-  \partial_{\bar{\theta}_-}. 
\eea
An appropriate representation for the gamma matrices is now given by
\bea 
\gamma^0 &=& \begin{pmatrix} 0 & 1 \\ 1 & 0 \end{pmatrix}, \\
\gamma^1 &=& \begin{pmatrix} 0 & -1 \\ 1 & 0 \end{pmatrix}, \\
C &=& D = \gamma^1, \\
\psi &=& \begin{pmatrix} \psi_- \\ \psi_+ \end{pmatrix}.
\eea
The operators $\{\text{L}_{+,n}, \text{Q}_{+,n}\}$ and $\{\text{L}_{-,n}, \text{Q}_{-,n}\}$ still form two independent copies of the super-Witt algebra; their commutation relations are still given by (\ref{Svira1}-\ref{Svira2}), with the only exception of the anticommutator of the "-" supercharges which is now:
\be
\{\text{Q}_{-,r}, \text{Q}_{-,s}\} =  -4 i \text{L}_{-,r+s}. \label{Svira_changed}
\ee{}
Although this change has mathematically speaking no importance, it should have one in the quantum theory. These consequences are beyond the scope of this article.





\section{Deformation of the action}

\subsection{The bosonic deformed action}

In order to present some aspect of the problem in a simpler way, let us first focus on the purely bosonic theory. The tensile bosonic string action is given by 
\be{} \label{Sbos}
S_{\text{Bos}} = - \frac{1}{4 \pi \ell} \int d^2 x \sqrt{-g} g^{\mu \nu} \partial_\mu X \partial_\nu X.
\ee{}
It is straightforward to see that the two possible deformations
\be{} \label{coord}
\text{(\textit{a})} \left\{ \begin{array}{c}
     t \mapsto \lambda t  \\
     x \mapsto x \\
     \ell \mapsto \frac{\ell}{\lambda} 
\end{array} \right. , \qquad
\text{(\textit{b})} \left\{ \begin{array}{c}
     t \mapsto t  \\
     x \mapsto \frac{x}{\lambda} \\
     \ell \mapsto \frac{\ell}{\lambda} 
\end{array} \right. ,
\ee{}
lead to the same result
\be{} \label{limSbos}
S_{\text{Bos}}(\lambda) = - \frac{1}{4 \pi \ell} \int dx dt (- \dot{X}^2 + \lambda^2 X'^2),
\ee{}
where we denote the derivative w.r.t to $t$ by a dot and the one w.r.t $x$ by a prime.
It is possible to, instead of deforming the coordinates like in (\ref{coord}), to deform the metric. Explicitly the deformation (\ref{coord}-\textit{a}) will be replaced by
\be{}
\left\{ \begin{array}{c}
     g^{-1} = \begin{pmatrix} -1 & \\ & 1 \end{pmatrix} \mapsto g^{-1} (\lambda) = \begin{pmatrix}-\frac{1}{\lambda^2} & \\ & 1 \end{pmatrix}   \\
     \ell \mapsto \frac{\ell}{\lambda} 
\end{array} \right. ,
\ee{}
and the deformation (\ref{coord}-\textit{b}) by
\be{}
\left\{ \begin{array}{c}
     g^{-1} = \begin{pmatrix} -1 & \\ & 1 \end{pmatrix} \mapsto g^{-1} (\lambda) = \begin{pmatrix}-1 & \\ & \lambda^2 \end{pmatrix}   \\
     \ell \mapsto \frac{\ell}{\lambda} 
\end{array} \right. .
\ee{}
Replacing in (\ref{Sbos}), we obtain again (\ref{limSbos}), showing that all four deformations are equivalent. A final remark on this equivalence: when the deformation $x \mapsto \frac{x}{\lambda}$ is chosen, the period of the (super-)strings $T$ has to be deformed as well by $T \mapsto \frac{T}{\lambda}$. Taking this into account, it is then straightforward to show that in both cases ($\{t \mapsto \lambda, T \mapsto T\}$ and $\{x \mapsto \frac{x}{\lambda}, T \mapsto \frac{T}{\lambda}\}$), the limit $\lambda \to 0$ of (\ref{mode_exp_of_X}) yields
\be \label{tensionless_X}
X = C_0 + \frac{2 \pi \ell}{T} P_0 t + \frac{i \sqrt{\ell}}{2 \pi} \sum_{n \in \mathds{Z} \setminus \{0\}} \frac{1}{n} \left(\alpha_n - \frac{2 \pi i n}{T}t \tilde{\alpha}_n \right) e^{- \frac{2 \pi i n x}{T}},
\ee 
with 
\be 
\alpha_n = \frac{1}{\sqrt{\lambda}}(a_n - \tilde{a}_{-n}), \qquad \tilde{\alpha}_n = \sqrt{\lambda}(a_n + \tilde{a}_{-n}). \label{tensionless_modes}
\ee 
in accordance with \cite{Ba1}.
 
\subsection{Deformation of the Clifford algebra}
Let us expose the strategy we will use. On the one hand, by analyzing the physical dimension, we understand that a deformation of the coordinates $x^\mu$ should be accompanied with a deformation of the spinor field $\psi$. On the other hand, a deformation of the metric has to be accompanied with a deformation of the Clifford algebra. Computing the deformation of the Clifford algebra from the deformation of the metric is quite easy, as we will show. Hence, in order to compute the deformation of the spinor, we will assume that the equivalence between the "coordinates deformation" point of view and the "metric deformation" point of view still hold, and compute the deformation of the spinor from the deformation of the gamma matrices. We remind that we use (\ref{exotic}) for the Clifford algebra.

\paragraph{} We now turn on computing the deformation of the Clifford algebra. We consider the case where the space-space component of the metric is deformed, which should be equivalent to $x \mapsto \frac{x}{\lambda}$. This will yield to a one-parameter family of Clifford algebras, denoted $\mathcal{C}(\lambda)$, whose generators $\Gamma^\mu (\lambda)$ satisfy 
\be 
\Gamma^\mu(\lambda) \Gamma^\nu(\lambda) + \Gamma^\nu(\lambda) \Gamma^\mu(\lambda) = -2 g^{\mu \nu} (\lambda) \mathds{1},
\ee
with 
\be{}
g^{\mu \nu} = \begin{pmatrix} -1 & \\ & \lambda^2 \end{pmatrix}.
\ee
It is important in this construction that we work with the inverse metric $g^{\mu\nu}$, and his associated gamma matrices "with upper indices". Indeed, the inverse metric converges, in that case, to a well defined degenerate matrix, whereas the normal metric diverges as $\lambda \to 0$. Had we consider the deformation equivalent to $t \mapsto \lambda t$, we would have done the opposite, i.e. consider the normal metric and gamma matrices "with lower indices".

\paragraph{}
The central problem the sought deformation should answer is to provide a representation of $\mathcal{C}(0)$ with a Majorana condition.  We will follow the strategy of \cite{Bulu}. For this, we see the full family of $\mathcal{C}(\lambda)$ as a collection of subalgebras of the bigger Clifford algebra Cl$(1,2)$, where the latter denotes the Clifford algebra associated to the three dimensional metric 
\be 
G_{\mu \nu} =  \begin{pmatrix} -1 & & \\ & 1 & \\ & & -1 \end{pmatrix}.
\ee
We will then use the fact that Cl$(1,2)$ admits a Majorana representation, and restrict the Majorana condition of the representation of Cl$(1,2)$ to the sub-representation of the $\mathcal{C}(\lambda)$. It is because of this injection of $\mathcal{C}(\lambda)$ into Cl$(1,2)$ that we have to work with the metric deformation equivalent to $x \mapsto \frac{x}{\lambda}$. If we would try instead the metric deformation equivalent to $t \mapsto \lambda t$, we should have seen the family $\mathcal{C}(\lambda)$ as family of subalgebras of Cl$(2,1)$, whose irreducible representations do not possess a Majorana condition, and we would not have obtained a reasonable Majorana condition for $\mathcal{C}(0)$.

\paragraph{} At this point, we have to ensure that the representation of Cl$(1,2)$ can be seen as an extension of the representation of $\mathcal{C}(1)$ we started with. This crucial point is doable only if we choose the Majorana conjugation matrix to satisfy $C^T = -C$. Indeed, we said that two kind of matrix, $C_+$, $C_-$ could be used in spacetime of dimension equal to 2, and this is no longer true in spacetime of dimension 3, where only $C_-$ exists. This is also what constrains us to embed our family $\mathcal{C}(\lambda)$ in a Clifford algebra of dimension 3. In higher dimension, the dimension of the representation will grow up, i.e. the spinors will have more components, which is an undesirable feature.

\paragraph{}
Explicitly,  let $\Gamma^0$, $\Gamma^1$ be the generators of $\mathcal{C}(1)$ (with identification $\Gamma^\mu \equiv \Gamma^\mu (1)$), with $\Gamma^0$ the timelike generator, $(\Gamma^0)^2 = \mathds{1}$, and $\Gamma_1$ the spacelike generator, $(\Gamma^1)^2 = -\mathds{1}$. Let $(V, \mathcal{R}_1)$ be an irreducible representation of the starting Clifford algebra $\mathcal{C}(1)$, used for example in \ref{action}, with $C$ and $D$ the matrices defining the Majorana and Dirac conjugation.. We have $\gamma^\mu = \mathcal{R}_1(\Gamma^\mu)$. Let $\Xi^0, \Xi^1$ and $\Xi^2$ be the generators of Cl$(1,2)$, $\Xi^1$ being the spacelike generator. Then $(V, \bar{\mathcal{R}})$ (with the same $V$), with $\bar{\mathcal{R}}(\Xi^0) = \gamma^0$, $\bar{\mathcal{R}}(\Xi^1) = \gamma^1$ and $\bar{\mathcal{R}}(\Xi^2) = \tilde{\gamma}1$ define an irreducible representation of Cl$(2,1)$. As argued in the previous paragraph, we chose the same matrices $C$ and $D$ to define the Majorana and Dirac conjugation.  An injection $\iota_\lambda : \mathcal{C}(\lambda) \to \text{Cl}(1,2)$ is given by
\bea{} \label{iota}
\iota_\lambda : &\Gamma^0(\lambda) \mapsto \Xi^0 ,& \\
&\Gamma^1(\lambda) \mapsto \frac{1 + \lambda^2}{2} \Xi^1 + \frac{1 - \lambda^2}{2} \Xi^2.&
\eea
We obtain a representation $(V,\mathcal{R}_\lambda)$ of $\mathcal{C}(\lambda)$ by putting $\mathcal{R}_\lambda = \mathcal{R} \circ \iota_\lambda$. The Majorana condition on any of these $\mathcal{R}_\lambda$ is obtained by restriction of the Majorana condition defined on $\bar{\mathcal{R}}$.

\subsection{Deformation of the spinor}

We finally compute the deformation of the spinors. The equivalence between the "metric deformation" and "coordinates deformation" stated before means that we are looking for a collection $\psi(\lambda)$ satisfying
\bea
\nonumber \mathcal{S}(\lambda) &=&
-\frac{1}{4\pi\ell(\lambda)}\int d^2 x(\lambda) \sqrt{-g} \big[ g^{\mu \nu} \partial_\mu(\lambda) X \partial_\mu(\lambda) X \\
&& \hspace{150pt} + i\bar{\psi}(\lambda) \gamma^\mu\partial_\nu(\lambda) \psi(\lambda)\big], \label{deformed_action} \\
&\doteq& -\frac{1}{4\pi\ell(\lambda)}\int d^2 x \sqrt{-g}(\lambda) \big[ g^{\mu \nu}(\lambda) \partial_\mu X \partial_\mu X + i\bar{\psi}  \gamma^\mu(\lambda)\partial_\nu \psi\big],
\eea
with the convention that for any quantity $\chi(\lambda)$ deformed in the l.h.s, its non-deformed counterpart in the r.h.s satisfies $\chi_{\text{r.h.s}} \doteq \chi(1)_{\text{l.h.s}}$; and reciprocally for quantities deformed in the r.h.s but not in the l.h.s.. Therefore, to compute the deformation of the spinor we first write the required equality:
\be{}
\bar{\psi}  \gamma^\mu(\lambda) \partial_\mu \psi =  \bar{\psi}(\lambda) \gamma^\mu \partial_\mu(\lambda) \psi(\lambda), \label{equivalenceofscaling}
\ee{}
leading to
\bea{}
\bar{\psi} \gamma^0 \partial_x \psi &=& \bar{\psi}(\lambda) \gamma^0 \partial_x \psi(\lambda), \\ 
\bar{\psi} \frac{\frac{1 + \lambda^2}{2}\gamma^0 + \frac{1 - \lambda^2}{2} \tilde{\gamma}}{\lambda} \partial_t \psi &=& \bar{\psi}(\lambda) \gamma^0 \partial_t \psi(\lambda).
\eea{}
We observe that, so long $\lambda \neq 0$, the two sets of matrices $\{ \gamma_0, \gamma_1 \}$ and $\{\frac{\frac{1 + \lambda^2}{2}\gamma_0 + \frac{1 - \lambda^2}{2} \tilde{\gamma}}{\lambda}, \gamma_1 \}$ obey the same Clifford relations. By uniqueness of the equivalence class of faithful irreducible representations of the Clifford algebras of even dimensions, we know that there exist an invertible matrix $P(\lambda)$ such that 
\bea{} \label{def_of_P_1}
P^{-1}(\lambda) \gamma^0 P(\lambda) &=& \gamma^0, \\
P^{-1}(\lambda) \gamma^1 P(\lambda) &=& \frac{\frac{1 + \lambda^2}{2}\gamma^1 + \frac{1 - \lambda^2}{2} \tilde{\gamma}}{\lambda}. \label{def_of_P_2}
\eea{}
Finally, the sought deformation is 
\be
\psi(\lambda) = P^{-1}(\lambda) \psi.
\ee
A last remark: in the previous section we said that the Majorana and Dirac conjugation matrices should be conserved in order to preserve the Majorana condition. This means that searching for $P(\lambda)$, we also need to consider the two following equations:
\bea
P^T(\lambda) C P(\lambda) &=& C, \label{def_of_P_3} \\
P^\dagger(\lambda) D P(\lambda) &=& D. \label{def_of_P_4}
\eea
We insist on the fact that the matrix $P(\lambda)$ is guaranteed to exist only if $\lambda \neq 0$. In the limit $\lambda \to 0$, this matrix might become singular. However, we expect the Lagrangian to have a non singular limit. 

\subsection{Explicit deformation and tensionless limit.}

We introduce a representation for the gamma matrices in which the equations (\ref{def_of_P_1} - \ref{def_of_P_4}) are easy to solve
\bea{}
\gamma^0 &=& \begin{pmatrix} 1 & 0 \\ 0 & -1 \end{pmatrix}, \\
\gamma^1 &=& \begin{pmatrix} 0 & 1 \\ -1 & 0 \end{pmatrix} = C = D.
\eea{}
The components of a spinor in this representation are $\psi = \begin{pmatrix} \psi_u \\ \psi_d \end{pmatrix}$. They are related to the $\psi_\pm$ by
\be 
\psi_u = \frac{1}{\sqrt{2}}(\psi_- + \psi_+), \qquad \psi_d = \frac{1}{\sqrt{2}} (\psi_- - \psi_+).
\ee 
In this representation
\be{} \label{updownP}
P(\lambda) = \begin{pmatrix} \sqrt{\lambda} & 0 \\ 0 & \frac{1}{\sqrt{\lambda}} \end{pmatrix},
\ee 
thus
\be{} \label{psi_lambda}
\psi(\lambda) = P^{-1}(\lambda)\psi = \begin{pmatrix} \frac{1}{\sqrt{\lambda}}\psi_u \\ \sqrt{\lambda}\psi_d \end{pmatrix}.
\ee{}
From here it is possible to take the limit $\lambda \to 0$ and obtain
\be{} \label{Stensionless}
\mathcal{S}_{\text{Tensionless}} = \frac{1}{4\pi \ell} \int d^2x \big[ \dot{X}^2  + i  \left(\psi_d \dot{\psi_u} + \psi_u \dot{\psi_d} +  \psi_u \psi_u' \right) \big].
\ee 
The mode expansion of the tensionless field $X$ has been already given in (\ref{tensionless_X}). For the spinor fields, we have seen that the well defined tensionless components are $\psi_u$ and $\psi_d$, whose mode expansions in the limit $\lambda \to 0$ are
\bea
&\psi_u[\lambda](x,t) \xrightarrow[\lambda \to 0]{} \chi(x) + t \tilde{\chi}'(x), \qquad \psi_d[\lambda](x,t) \xrightarrow[\lambda \to 0]{} \tilde{\chi}(x),& \\
&\chi(x) =\sqrt{2 \pi \ell} \sum_n \beta_n e^{-\frac{2 \pi i n x}{T}}, \qquad \tilde{\chi}(x) = \sqrt{2 \pi \ell} \sum_n \tilde{\beta}_n e^{-\frac{2 \pi i n x}{T}},
\eea 
with
\be 
\beta_n = \frac{b_n + \tilde{b}_{-n}}{\lambda}, \qquad \tilde{\beta}_n = b_n - \tilde{b}_{-n}. \label{deformation_of_coef}
\ee 

Note that it is possible to play exactly the same game starting from the standard super-string action, in which case the tensionless limit is
\be
\mathcal{S}_{\text{Alternative}} = \frac{1}{4\pi \ell} \int d^2x \big[ \dot{X}^2  + i  \psi_d \dot{\psi_d}  \big]. 
\ee 
This result explains why we have to start with the exotic action.

\subsection{Deformation of the symmetries}

For $\lambda \neq 0$, the symmetries of the deformed action (\ref{deformed_action}) are just the expressions L$_\pm$, Q$_\pm$ given in sections 1 and 2, but with $\lambda$ dependent quantities. For example
\bea 
\text{L}_+[\lambda](\xi_+(\lambda)) &=& \xi_+^\mu(\lambda) \partial_{+\mu}(\lambda) + \frac{1}{2} \partial_{+\mu}(\lambda) \xi^\mu_+(\lambda)   \bar{\theta}_+(\lambda) \partial_{\bar{\theta}_+}(\lambda).
\eea
Here the $\lambda$-dependence of $\xi_\pm$ is just through $x^\pm$ ($\xi_+(\lambda) \doteq \xi_+ (x^+(\lambda))$), whereas $\theta_\pm$ need to be changed like $\psi$ using the formula (\ref{psi_lambda}). However, at $\lambda = 0$, the $\pm$ decomposition of the symmetries do not hold anymore and is replaced by another decomposition. In other words, we have the equalities
\bea
\text{L}[\lambda](\xi(\lambda)) &=& \text{L}_+[\lambda](\xi_+(\lambda)) + \text{L}_-[\lambda](\xi_-(\lambda)) = \text{K}(f) + \text{M}(g) + o(\lambda), \\
\text{Q}[\lambda](\bar{\epsilon}(\lambda)) &=& \text{Q}[\lambda](\bar{\epsilon}(\lambda)) + \text{Q}[\lambda](\bar{\epsilon}(\lambda)) =  \text{G}(\bar{\zeta}) + \text{H}(\bar{\rho}) + o(\lambda),
\eea 
but it is impossible to express for example K$(f)$ alone in terms of L$_+$ and L$_-$. It means that the decomposition in term of projectors $h_\pm$, P$_\pm$ do not exist in the tensionless limits; or equivalently, these projectors do not possess a well defined limit when $\lambda$ goes to $0$. The new symmetries are
\bea 
\text{M}(g) &=& g \partial_t + \frac{1}{2} g' \bar{\theta_u} \partial_{\bar{\theta}_d}, \\
\text{K}(f) &=& f \partial_x + \frac{1}{2} f' (\bar{\theta}_u \partial_{\bar{\theta}_u} + \bar{\theta}_d \partial_{\bar{\theta}_d}) + t\text{M}(f'), \\
\text{H}(\bar{\rho}) &=& \bar{\rho}(i\partial_{\bar{\theta}_d} -  \bar{\theta}_u \partial_t),  \label{def_of_H} \\
\text{G}(\bar{\zeta}) &=&  \bar{\zeta}( i \partial_{\bar{\theta}_u} + \bar{\theta}_u \partial_x - \bar{\theta}_d \partial_t) - t\text{H}(\bar{\zeta}'). \label{def_of_G} \\
\eea
and $f,g, \zeta, \rho$ are related to $\xi$ and $\epsilon$ by
\bea 
\frac{\xi^+ - \xi^-}{2\lambda} = f, &\qquad&  \frac{\xi^+ + \xi^-}{2} = g + tf' \\
\frac{\bar{\epsilon}_u}{\sqrt{\lambda}} = \bar{\zeta}, &\qquad& \sqrt{\lambda}\bar{\epsilon}_d = \bar{\rho} - t\bar{\zeta}'.
\eea 





Although the symmetries L$_\pm$, Q$_\pm$ and K, M, G, H are not directly related, their modes are, by
\bea{} \label{contra1}
\text{K}_{n} = \frac{ \text{L}_{+,n} - \text{L}_{-,-n} }{2 \lambda}, \qquad \text{M}_{n} = \frac{ \text{L}_{+,n} - \text{L}_{-,-n} }{2}, \\
\text{G}_{r} = \frac{1}{\sqrt{2\lambda}}(\text{Q}_{-,r} - \text{Q}_{+,-r}), \qquad \text{H}_{r} = \sqrt{\frac{\lambda}{2}}(\text{Q}_{+,r} + \text{Q}_{-,-r}), \label{contra2}
\eea{} 
which is a (disguised) Wigner-Inönü contraction. Why the Wigner-Inönü contraction takes this form is understood by looking at (\ref{Svira1}-\ref{Svira2}), we see that $T$ appears in the structure constants. As said earlier, the deformation $x \mapsto \frac{x}{\lambda}$ implies a deformation $T \mapsto \frac{T}{\lambda}$ and thus we get a $\lambda$-dependent algebra. In order to remove this $\lambda$-dependence, it is possible to scale L$_{\pm,n} \mapsto \frac{1}{\lambda} \text{L}_{\pm,n}$, Q$_{\pm,r} \mapsto \frac{1}{\sqrt{\lambda}}\text{Q}_{\pm,r}$, after what equations (\ref{contra1}-\ref{contra2}) take the form of a standard Wigner-Inönü contraction. The non-vanishing commutation relations of the new symmetry modes are
\bea{}
[\text{K}_{n}, \text{K}_{m}] &=& \frac{2 \pi}{T}i(m-n) \text{K}_{n+m}, \\
{}[\text{K}_{n}, \text{M}_{m}] &=& \frac{2 \pi}{T}i(m-n) \text{M}_{n+m}, \\
{}[\text{K}_{n}, \text{G}_{r}] &=& \frac{2 \pi i}{T}(r-\frac{n}{2})\text{G}_{n+r}, \\
{}[\text{K}_{n}, \text{H}_{r}] &=& \frac{2 \pi i}{T}(r-\frac{n}{2})\text{H}_{n+r}, \\
{}[\text{M}_{n}, \text{G}_{r}] &=& \frac{2 \pi i}{T}(r-\frac{n}{2}) \text{H}_{n+r}, \\
\{\text{G}_{r}, \text{G}_{s}\} &=& 2 i \text{K}_{r+s}, \\
\{\text{G}_{r}, \text{H}_{s} \} &=& 2 i \text{M}_{r+s}.
\eea{}
which are recognized as the commutation relations of the super-BMS$_3$ algebra \cite{AGS}. The fact that the algebra of symmetries of the tensionless action (\ref{Stensionless}) is the super-BMS$_3$ algebra, as well as its relation with the algebra of symmetries of the tensile action, was already shown in \cite{Ba3}. Thus, what this last paragraph shows is that the implementation of the Majorana condition haven't altered the previous results of \cite{Ba1}, \cite{Ba2} and \cite{Ba3}. Furthermore, we can now state that the generators G and H defined in (\ref{def_of_H}-\ref{def_of_G}) are real.

\section{Conclusion}

In this article, we have successfully given a Majorana condition for the spinors in the tensionless limit of the super-Polyakov action. This Majorana condition is obtained by carefully looking at the deformation linking the exotic super-Polyakov action to its tensionless limit; in particular we have computed the deformation of the spinor fields by preserving the equivalence between a deformation of the coordinates and a deformation of the metric. This deformation has also been constructed in a way that it preserves the Majorana condition existing in the tensile theory. It was not guaranteed at all that such deformation was possible, and a profound analysis of representation theory of Clifford algebras has been done to ensure its existence. It has been shown that it was crucial at this step to deform the spacelike coordinate $x$ and not the timelike coordinate $t$, partially explaining why all previous attempts of defining a Majorana condition in the tensionless limit failed. We have also make sure that the newly computed deformation reproduce some of the most important results shown previously concerning the tensionless action. The way on how symmetries are deformed has been our main focus regarding this point; and we correctly obtain the Wigner-Inönü deformation of the two copies of the super-Witt algebra to the super-BMS$_3$ algebra. As a side result, we have pointed out the little discrepancies existing between the symmetry of the standard super-Polyakov action and the exotic one. These discrepancies are significant only at the quantum level and should be explored in the future.  

\subsection*{Acknowledgment}

I would like to thanks my supervisor Mokhtar Hassaine for his precious comments.

\appendix{}

\section*{Appendix}

In this appendix we give the explicit representation of the symmetry modes in term of  super-differential operators. 

\paragraph{Symmetry modes of the standard super-Polyakov action}
\bea
\text{L}_{+,n} &=& e^{-\frac{2 \pi i n x^+}{T}}  \left( \partial_+ - \frac{\pi i n}{T} \bar{\theta}_+ \partial_{\bar{\theta}_+} \right), \label{L_+_n}\\
\text{L}_{-,n} &=& e^{-\frac{2 \pi i n x^-}{T}} \left( \partial_- - \frac{\pi i n}{T} \bar{\theta}_- \partial_{\bar{\theta}_-} \right), \label{L_-_n}  \\
\text{Q}_{+,n} &=& e^{-\frac{2 \pi i n x^+}{T}} \left( -i \partial_{\bar{\theta}_+} - 2 \bar{\theta}_+ \partial_+   \right), \\
\text{Q}_{-,n} &=& e^{-\frac{2 \pi i n x^-}{T}} \left( -i \partial_{\bar{\theta}_-} - 2 \bar{\theta}_- \partial_-   \right).
\eea

\paragraph{Fermionic symmetry modes of the exotic super-Polyakov action} (The bosonic modes are indentical to those of the standard super-Polyakov action).
\bea
\text{Q}_{+,n} &=& e^{-\frac{2 \pi i n x^+}{T}} \left(i \partial_{\bar{\theta}_+} + 2 \bar{\theta}_+ \partial_+   \right), \\
\text{Q}_{-,n} &=& e^{-\frac{2 \pi i n x^-}{T}} \left( i \partial_{\bar{\theta}_-} - 2 \bar{\theta}_- \partial_-   \right).
\eea

\paragraph{Symmetry modes of the tensionless action}
\bea{}
\text{K}_{n} &=&  e^{-\frac{2 \pi i n x}{T}} (\text{I}_n - \frac{2 \pi i n t}{T}  \text{J}_n), \\
\text{M}_{n} &=&  e^{-\frac{ 2 \pi i n x}{T}} \text{J}_n, \\
\text{G}_{r} &=& e^{-\frac{ 2 \pi ir x}{T} } (\text{U} + \frac{2 \pi i r t}{T}  \text{V} ),\\
\text{H}_{r} &=& e^{-\frac{ 2 \pi ir x}{T}} \text{V},
\eea 
with
\bea{}
\text{I}_n &=& \partial_x - \frac{i \pi n}{T}(\bar{\theta}_u  \partial_{\bar{\theta}_u} + \bar{\theta}_d \partial_{\bar{\theta}_d}), \\
\text{J}_n &=& \partial_t - \frac{i \pi n}{T} \bar{\theta}_d \partial_{\bar{\theta}_u},\\
\text{U} &=&  i \partial_{\bar{\theta}_d} + \bar{\theta}_d \partial_x - \bar{\theta}_u \partial_t, \\
\text{V} &=& i\partial_{\bar{\theta}_u} -  \bar{\theta}_d \partial_t .
\eea

\printbibliography

\end{document}